# Electronically reconfigurable metal-on-silicon metamaterial

Yaroslav Urzhumov[1*], Jae Seung Lee[2], Talmage Tyler[1], Sulochana Dhar[1], Vinh Nguyen[1],

Nan M. Jokerst[1], Paul Schmalenberg[2] & David R. Smith[1]

[1]Center for Metamaterials and Integrated Plasmonics, Duke University, Durham, North Carolina 27708, USA.

[2]Toyota Research Institute of North America, Ann Arbor, Michigan 48105, USA

ABSTRACT

Reconfigurable metamaterial-based apertures can play a unique role in both imaging and in beam-forming applications, where current technology relies mostly on the fabrication and integration of large detector or antenna arrays. Here we report the experimental demonstration of a voltage-controlled, silicon-based electromagnetic metamaterial operating in the W-band (75-110 GHz). In this composite semiconductor metamaterial, patterned gold metamaterial elements serve both to manage electromagnetic wave propagation while simultaneously acting as electrical Schottky contacts that control the local conductivity of the semiconductor substrate. The active device layers consist of a patterned metal on 2 μm thick *n*-doped silicon layer, adhesively bonded to a transparent Pyrex wafer. The transmittance of the composite metamaterial can be modulated over a given frequency band as a function of bias voltage. We demonstrate a quantitative understanding of the composite device through the application of numerical approaches that simultaneously treat the semiconductor junction physics as well as wave propagation.

# 1. Introduction: Metal-semiconductor reconfigurable metamaterials

Metamaterials are artificial media whose properties derive primarily from structure rather than composition[1-3]. First proposed as a means of achieving enhanced electromagnetic response[4], metamaterials have been used to demonstrate unprecedented material properties, such as negative refractive index[5-7], as well as unconventional devices such as invisibility cloaks[8-10]. Yet, in spite of their successes, passive, linear metamaterials represent only a first step in harnessing the advantages of artificial media. Metamaterials are, in fact, ideally suited for the development of active and dynamically tunable materials that offer enormous opportunities that span multiple technologies, including beam-forming and imaging. Because the local electric and magnetic field distributions within and around metamaterial inclusions are strongly inhomogeneous, the local field amplitudes can be orders of magnitude larger than that of the incident field. This field enhancement is especially large near the capacitive regions associated with metallic inclusions. A material in the vicinity of these large field regions will have a disproportionate influence on the effective electromagnetic properties of the composite; for instance, a modest change in the dielectric properties of an embedded material, however achieved, can result in much larger variation of the effective constitutive parameters of the composite.

Naturally occurring materials exist whose electromagnetic properties can be manipulated with various external stimuli, including light, temperature, strain, magnetic field and voltage. As a result, tunable hybrid metamaterials have been demonstrated that make use of a materials such as ferrites[11], superconductors[12], vanadium dioxide,[13,14] strontium titanate[15] and graphene[16] integrated into metamaterial elements. Other physical mechanisms have also been proposed for introducing reconfigurability, including mechanical changes and deformations to either the metamaterial inclusions or their local environment[17-20]. All materials that can be controlled via external modulation are of interest, since even the weakest response can be substantially magnified through interaction with metamaterial inclusions.

One class of materials particularly attractive for integration into reconfigurable metamaterials is semiconductors. The conductivity of a semiconductor can be altered by a variety of mechanisms, including applied voltage or illumination by light. Semiconductor fabrication is mature, with methods for dense integration of active devices now well established. In 2006, Padilla *et al.*[21] applied photodoping to modulate the carrier concentration in a high resistivity gallium arsenide substrate, on top of which a copper metamaterial structure was patterned by photolithographic techniques. The metamaterial in this experiment was designed to exhibit a tunable transmittance at terahertz frequencies. As a function of the incident light intensity, the conductivity of the substrate increased, to the point that the metamaterial response could be damped; the composite structure thus behaved as a light-activated switch for terahertz radiation.

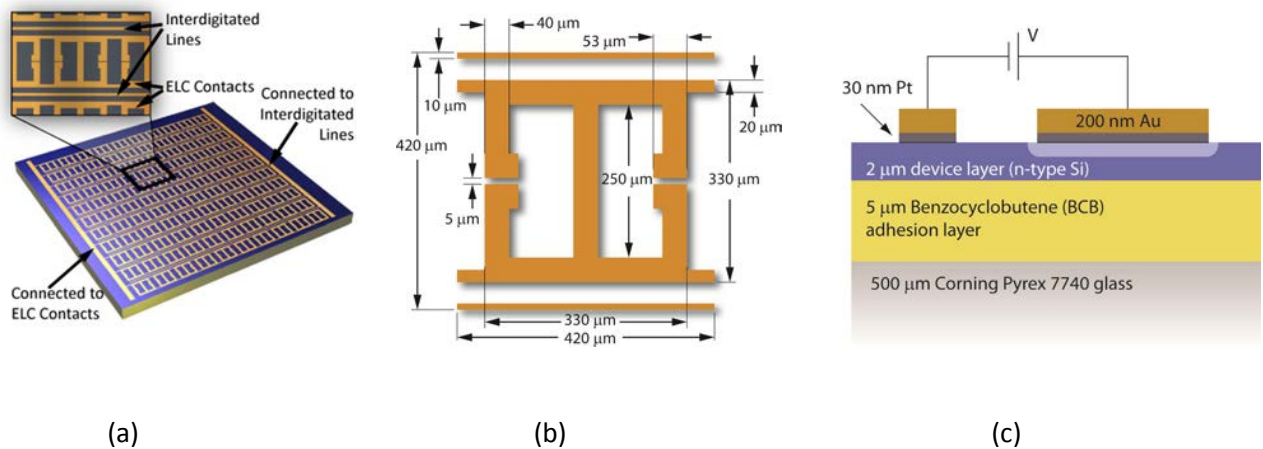

(a)　　　　　　　　　　　　　(b)　　　　　　　　　　　　　(c)

**Figure 1 | Silicon-ELC metamaterial.** (**a**) Depiction of the planar structure, which consists of a layer of ELC metamaterial elements deposited on a 2 μm thick wafer of n-doped Si, bonded to a Pyrex glass substrate. (**b**) Dimensions of a single ELC element and bias lines. (**c**) Side view schematic, showing the various fabrication layers. Note the depletion region around the reverse biased Pt/Au Schottky contact.

While photodoping provides an intriguing means of introducing reconfigurability[21-25]—potentially avoiding the layout of an electrical interconnection structure—the required power levels to achieve significant modulation of the signal are not necessarily practical for devices at all regions of the spectrum. As an alternative approach, the carrier concentration within a semiconductor can be varied via the application of a voltage across an appropriate metal-semiconductor junction[26-29]. For metamaterials based on metallic inclusions, portions of each metamaterial element can serve as a contact—forming a junction with a semiconductor substrate material—while the composite element scatters radiation according to the metamaterial design. A key example of a voltage-controlled metamaterial was reported by Chen *et al.*[26], who used a metamaterial layer to form a Schottky diode with a 3 μm layer of *n*-doped gallium arsenide. By applying a reverse bias voltage to the metamaterial elements, the width of the depletion region at the metamaterial-semiconductor interface was increased, reducing the local conductivity and changing the effective metamaterial response. Voltage controlled semiconductor-based metamaterials have now been demonstrated for several potential applications[26-28], with switching speeds of up to 10 MHz for a metamaterial operating at 0.5 THz having been demonstrated[28,29].

A large number of semiconducting materials exist, all of which can potentially serve as the tunable components within structured metamaterial inclusions. However, the substantial industry that has developed around silicon technology makes silicon an ideal starting point for more complex, semiconductor-based, reconfigurable metamaterials. Fabrication techniques are abundant and well-developed, as is the infrastructure for the production of silicon (Si) devices. Although Si device switching speeds are not as inherently fast as those based on gallium arsenide (GaAs), larger wafer areas and lower material costs make Si-based reconfigurable metamaterials attractive for a wide range of sub-terahertz imaging and radar devices.

In the present work, we propose, fabricate and test a reconfigurable sub-terahertz metamaterial based on a Schottky diode formed between platinum metal and single crystal thin film Si, operating at W-band frequencies (75-110 GHz). The W-band portion of the electromagnetic spectrum is used for satellite communications and millimeter wave radars, with a sub-band centered around 77 GHz being allocated and used for automotive collision avoidance radar. Our device is formed using thin film Si from silicon-on-insulator (SOI), bonded to a Pyrex substrate to form a thin film active layer on a transparent substrate, thus minimizing absorption losses. Controlling the loss in metamaterial structures is critical; the use of thick semiconductor substrates in the sub-terahertz and terahertz bands can result in losses due to substrate absorption. Since the active thickness of the semiconductor used in the metamaterials demonstrated here is on the order of a micron, the losses associated with a thicker semiconductor substrate (typically 350-500 microns thick) are avoided.

To aid in the design and interpretation of the composite metamaterial, we apply a finite-element based numerical approach for self-consistent, integrated modeling of electromagnetic and electronic processes in metal-semiconductor metamaterials. This combined electronic-electromagnetic design represents a key step towards the manufacture of tunable metamaterial devices that optimally leverage semiconductor physics.

## 2. Metal-on-Silicon metamaterial design

The tunable metamaterial presented herein is based on a passive metamaterial layer, for which we make use of an electrically coupled inductive-capacitive (ELC) resonator[30] similar to that demonstrated at terahertz frequencies by Padilla *et al.*[21]. The ELC is a resonant inclusion that couples to the electric component of an incident electromagnetic wave. The ELC metamaterial is convenient in that a single, planar layer can be fabricated and tested; a magnetic metamaterial would require some depth in the propagation direction and would not be as simple to fabricate using commercial lithographic processing. An ELC design with two parallel capacitive gaps rather than a single gap is used to achieve a slightly more compact inclusion (Fig. 1b).

Reconfigurability can be introduced into the otherwise passive metamaterial by having the ELC metamaterial inclusions serve simultaneously as metal contacts to semiconductor devices. For the structure presented here, a Schottky diode is formed at the interface between the ELC and a semiconductor layer. Electrical contacts must be introduced to apply a voltage bias to the ELCs (and hence across the Schottky barrier), while at the same time not interfering with the desired electromagnetic response. Orienting the conducting lines along a direction perpendicular to that of the electric field of the incident wave eliminates unwanted scattering, and thus the ELC inclusions can be electrically connected in the manner shown in Fig. 1.

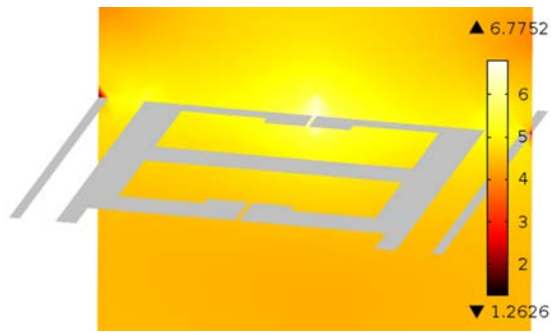

**Figure 2 | Simulated fields for the Si-ELC metamaterial.** Plot showing field strength on a plane that cuts through the ELC. Finite element method frequency-domain simulation performed near resonance at 75 GHz, with zero dc bias.

To form the integrated devices, Schottky contact metallic ELCs are deposited and patterned on a thin film layer of *n*-doped Si bonded to a translucent Pyrex substrate. The same Schottky metallization is also used to simultaneously pattern lines interdigitated between the rows of metamaterial elements, which serve as a second contact. Grounding the ELCs and applying a positive voltage to the interdigitated lines creates a forward bias at the metal/semiconductor Schottky interface, while simultaneously reverse biasing the ELC Schottky contacts and creating a depletion region just below the ELCs. Since only a variable conductivity is desired here, and static currents are unnecessary, this design avoids creating a separately deposited Ohmic contact, which would add complexity to the fabrication.

With its large dielectric constant ($\varepsilon_{Si} \approx 11.8$), Si presents a significant impedance mismatch to air at W-band frequencies. Even in the absence of doping, a sample with a thick Si substrate would be highly reflective. To minimize the impedance mismatch, the ELC metamaterial is fabricated on a thin (2 μm) layer of *n*-type silicon that is bonded to a 500 μm thick Pyrex 7740 glass wafer. Pyrex is a low loss insulator at millimeter wave frequencies ($\varepsilon_{Py} \approx 4.2$ at W-band), while the doped silicon layer is thin enough to be minimally absorptive.

The permittivity of the ELC metamaterial was designed to exhibit a resonance within the 60-90 GHz frequency band. For the initial design, a full-wave commercial finite-element solver (COMSOL Multiphysics, details below) was used to simulate the scattering from the ELC metamaterial. The Si conductivity and frequency dispersion were ignored in these first simulations, approximating the fully depleted condition. Field profiles in and around the ELC are presented in Fig. 2. While the majority of the local field is spread out over a region of 1-2 mm around the ELC, there are highly localized fields that occur directly in the capacitive gap region. These fields are likely to interact most strongly with the 2 μm Si layer directly below.

The combination of the resonant ELC metamaterial and a positive-ε dielectric substrate can result in a frequency band where the composite is roughly matched to free space. This matching occurs in the region where the negative permittivity of the ELCs offsets the positive permittivity of the substrate, and is characterized by a minimum in the reflectance. For the initial simulations, a Pyrex thickness of 175 μm was used so that this impedance matched region could be studied. In the experiments, 175 μm thick Pyrex wafers were initially used, but proved too fragile, so that 500 μm wafers were ultimately used in the experiments and subsequent simulations. We adjusted the various geometrical dimensions of the ELCs to ensure that both the resonant frequency of the ELCs (indicated by a dip in the transmission, or $S_{21}$) and the free space impedance match frequency (indicated by a dip in the reflection, or $S_{11}$) could be observed over the 60-90 GHz frequency band. While adjusting dimensions to optimize the scattering characteristics of the ELC metamaterial, a minimum 5 μm feature size was assumed. The fabricated samples had the ELC gap size close to 4 μm, as observed in the optical microphotographs. From these simplified simulations we arrived at the ELC dimensions shown in Fig. 1b. Based on the location of the dip in the magnitude of the $S_{21}$ response (not shown), these ELCs have resonances approximately at 70 GHz.

Both the resonance and impedance match frequencies are of potential interest for active or tunable devices. Because of the need for the larger substrate, however, we could only probe frequencies near the resonance to observe modulation introduced by applying a bias voltage.

For an initial estimate of an appropriate doping level for the *n*-type silicon layer, simulations were performed as described below to ascertain the amount of doping needed to fully deplete the charge carriers in the region beneath the ELC capacitive gaps under a reverse bias well below the breakdown voltage. A doping level of $(1-2) \times 10^{15} \, \text{cm}^{-3}$ theoretically enables full vertical depletion at a reverse bias voltage of 5-7 V, assuming Si thickness of 2 μm. The reverse bias on the Schottky diode depletes the Si semiconductor material near the metal-semiconductor interface. A reverse bias voltage of 7 V would vertically fully deplete the Si directly beneath the Schottky metallization in the ELC. Application of additional reverse bias would deplete laterally away from the metal-semiconductor junction, thus further increasing the depletion region in the ELC gap.

### 3. Fabrication methodology and procedure

The Schottky diodes, both for the reverse biased ELCs and for the forward biased interdigitated contacts, were deposited onto a layer of Si bonded to Pyrex. The process began with a silicon-on-insulator (SOI) wafer, which was purchased from Addison Engineering Inc. (San Jose, CA) in the form of a 100 mm-diameter wafer. The *n*-type phosphorus-doped (2±0.5) μm-thick monocrystalline silicon layer (<100> orientation) came on a 1 μm buried oxide (BOX) $SiO_2$ layer on a 400 μm-thick Si handle wafer (<100> orientation). The Si device layer conductivity specification by the manufacturer was (1÷20) Ω-cm, which was validated by a four point probe measurement of $\sigma=(2\pm0.2)$ $\Omega$-cm, corresponding to a doping of $N_d=(2.3\pm0.2) \times 10^{15}$ cm$^{-3}$ under the assumption that the two are related by the simple formula[31] $\sigma=e\ \mu_e(N_d)\ N_d$, where $\mu_e(N_d)=0.132$ m$^2$/(V·s) at the expected doping level.

Before testing the metamaterial sample, the Schottky diode characteristic of the Si device layer was first measured to test the quality of the junctions. Pt/Au (30 nm/200 nm) Schottky contact characterization pads (500 μm x 500 μm square pads separated by 10 μm) were deposited onto the top surface of the Si device layer. The current-voltage (I-V) characteristic measured from these pads, presented in Fig. 3a, shows a breakdown voltage of approximately 25 V. The breakdown voltage of reverse biased Schottky diodes is dominated by the material quality and the surface area of the metal-semiconductor contacts. As the surface area of Schottky diodes increases, the reverse bias breakdown voltage decreases due to material and nanofabrication imperfections.

The final ELC metamaterial sample was fabricated on a 4 inch SOI wafer, with a total effective surface area of the Schottky ELC of 1,050 mm$^2$, resulting in the I-V curve shown in Fig. 3b. The I-V curve indicates a soft reverse bias breakdown. Since the material quality for the Si device layer in SOI can vary based upon the SOI fabrication process, two fabrication approaches were attempted, with the large area Schottky contacts deposited (1) on the top of the Si device layer while it was still part of the SOI wafer; and (2) on the oxide side of the SOI wafer, after the Si device layer was separated from the SOI. Both I-V characteristics exhibited the soft reverse bias breakdown shown in Fig. 3b. Thus, the Si device layer quality was found to be similar on both sides.

The fabrication process proceeded as follows. Thermal-compressive bonding was used to bond the SOI wafer to a 100 mm diameter borosilicate glass wafer (Pyrex® 7740, 500 μm thick), using a 5 μm layer of benzocyclobutene (BCB, DOW) as an adhesive bonding layer. The 400 μm-thick SOI handle wafer was removed chemically in a heated KOH bath; the remaining KOH-resistant

BOX layer was then removed with buffered oxide etch (BOE), leaving only the 2 μm device layer (Si) bonded to the glass wafer. The ELCs and interdigitated electrodes were simultaneously patterned on the device layer using standard photolithography and lift-off. The metal stack was electron-beam evaporated, and consisted of a 30 nm Pt Schottky contact and 200 nm Au capping layer. The completed device is depicted in Fig. 1c.

**4. Integrated modeling platform for coupled electronic-electromagnetic processes in semiconductors**

To obtain a quantitative description of the voltage dependent behavior of the Si-ELC metamaterial, the distribution of the conductivity as a function of bias voltage was computed using COMSOL Multiphysics[32]. To a good approximation, the conduction electrons and holes in a semiconductor crystal can be modeled as a two component plasma.[33] Electromagnetic waves with frequencies in the terahertz range and below do not have enough energy to generate photo-excited electrons, or to probe interband and atomic transitions in a wide-gap semiconductor such as Si. The dielectric function of Si can be thus approximated accurately by the Drude formula for a two-fluid plasma[34]:

$$\varepsilon_{Si}(\omega) = \varepsilon_b - \frac{\omega_{pe}^2}{\omega(\omega + i\gamma_e)} - \frac{\omega_{ph}^2}{\omega(\omega + i\gamma_h)}. \tag{1}$$

Here, $\varepsilon_b \approx 11.8$ is the relative permittivity due to the host lattice, $\omega_{pe,h} = (e^2 n_{e,h} / \varepsilon_0 m^*_{e,h})^{1/2}$ is the plasma frequency for conduction electrons (holes) of number density $n_e$ ($n_h$) and effective mass $m^*_e$ ($m^*_h$), and $\gamma_{e,h}$ are the electron and hole collision frequencies. Here and in what follows, $e > 0$ is the positive elementary charge constant; the charge carried by an electron is thus $q_e = -e$ and a hole carries $q_h = e$. Considering strongly $n$-doped silicon, we can assume that $n_h \ll n_e$ and $\omega_{ph} \ll \omega_{pe}$, and can therefore neglect the hole contribution to the dielectric function. In what follows, we use the effective electron mass in Si$^{35}$ $m_e^* = 1.08\, m_e$ and a damping rate of $\gamma_e = \left(2.2 \times 10^{-13}\, s\right)^{-1}$.

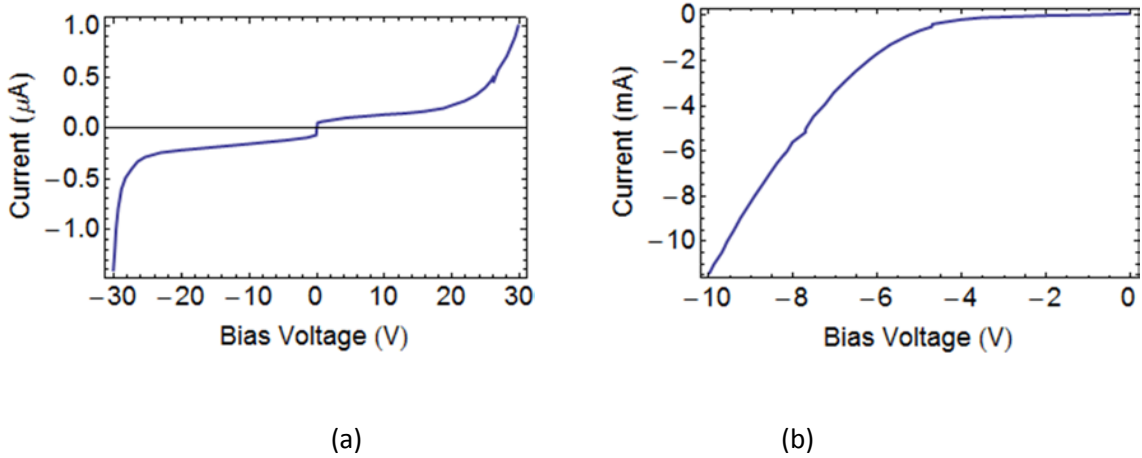

(a)                                                                          (b)

**Figure 3 | Measured I-V curves of Si structures.** (**a**) I-V characteristic of the Si device layer on the SOI wafer using 500 μm x 500 μm pads separated by 10 μm. (**b**) I-V characteristic of the Schottky diode formed on the Si device layer surface facing the SOI oxide layer. The Schottky metallization was deposited after the Si device layer has been bonded to a Pyrex host substrate. The total surface area of the reverse biased ELC Schottky contact is approximately 1,050 mm$^2$.

Our goal is to modify the dielectric function of Si by inducing a change in the electron concentration $n_e$, which is accomplished by the application of an external quasistatic electric field. Before considering the full numerical simulations, it is useful to note that the application of a bias voltage in the present scenario leads to the modulation of the *depletion region* at the interface between the metal and semiconductor. The standard equilibrium Schottky band diagram is illustrated in Fig. 4a, showing the barrier produced when the Schottky contact is formed[31].

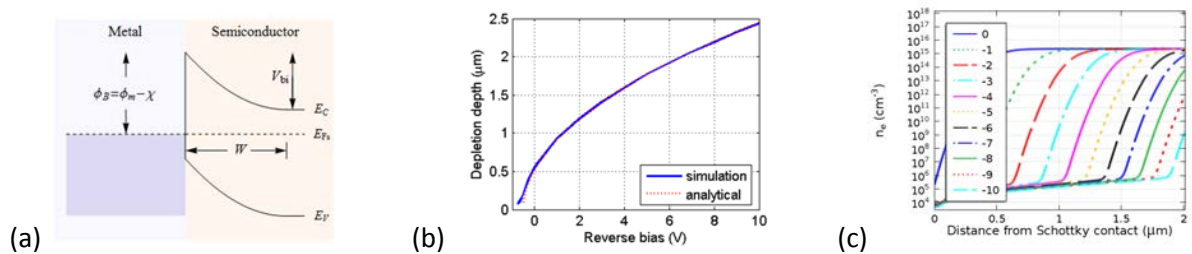

**Figure 4 | Simulated 1D carrier profiles for a silicon Schottky barrier.** (**a**) Standard band diagram of an idealized Schottky contact. (**b**) Depletion depth versus reverse bias voltage for the Schottky barrier used in the metamaterial. (**c**) Electron concentration (in cm$^{-3}$, log scale) distribution in silicon as a function of the distance from the Schottky contact, for various reverse bias voltages indicated by the legend.

At the Schottky interface the voltage in the semiconductor differs from that of the metal by a finite value known as the barrier height $\phi_B$. Under the assumptions of a perfect gapless contact, the barrier height $\phi_B$ is simply the difference between the electron work function of the metal $\phi_m$ (the energy difference between the Fermi level and vacuum level), and the electron affinity in the semiconductor $\chi_{semi}$ ($\approx 4\,eV$ for silicon[35,36]), measured from the bottom of the conduction band, that is, $\phi_B = \phi_m - \chi_{semi}$. For Si/Pt contacts[36-38], the barrier height is $\phi_B \approx 0.83\,eV$. The potential drop experienced by a charge carrier moved across the depletion region is the *built-in* potential, $V_{bi}$, which has the form $V_{bi} = \phi_m - \chi_{semi} - \phi_n$, where $\phi_n$ can be estimated from $\phi_n = E_g/2 - kT\ln(n_e/n_i) \simeq 0.312\,eV$. The depth of the depletion region can thus be estimated with the formula

$$W = \sqrt{\frac{2\varepsilon(V_{bi}+V)}{en_e}}, \qquad (2)$$

which assumes a perfect one-dimensional interface and a reverse bias voltage $V$. Eq. (2), plotted in Fig. 4b, predicts a depletion depth of ~0.5 μm in the absence of a bias voltage, increasing to approximately 2.4 μm with a 10 V bias. The depletion depth is relatively void of charge carriers and thus the region immediately adjacent to the metamaterial changes character from conducting to nearly insulating. Here, it becomes immediately clear that modulating the Schottky contact bias voltage extends the insulating region and pushes the lossy semiconductor region further away from the metamaterial elements.

Given the nature of the electromagnetic near-fields around the metamaterial inclusions, which are on the millimeter scale (Fig. 2), the altered capacitance associated with the Schottky barrier can hardly be expected to produce a shift in the resonance frequency of the metamaterial. However, the change in conductivity can be expected to influence the damping of the ELC resonance, so that the semiconductor layer thus behaves as a controllable thin-film absorber. Indeed, at the W-band frequencies of interest, $\omega \ll \gamma_e$ so that we may write Eq. (1) to good approximation as

$$\varepsilon_{Si} \simeq \left(\varepsilon_b - \frac{\omega_{pe}^2}{\gamma_e^2}\right) + \frac{i\omega_{pe}^2}{\omega\gamma_e} \simeq \frac{i\omega_{pe}^2}{\omega\gamma_e}. \qquad (3)$$

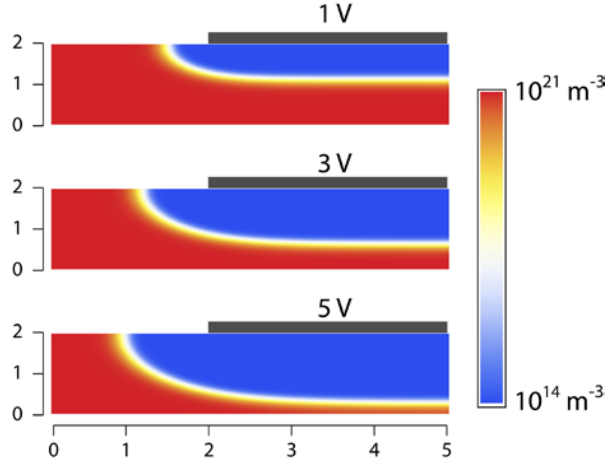

**Figure 5 | 2D Simulations of carrier density profile.** As reverse bias increases, the depletion layer increases below the ELC electrode. (**a**) COMSOL 2D simulations of single Schottky diode, for *n*-type Si with doping 2.3·10$^{15}$ cm$^{-3}$, 1 V reverse bias; (**b**) 3 V reverse bias; (**c**) 5 V reverse bias.

To better integrate the semiconductor and electromagnetic modeling steps, we make use here of COMSOL Multiphysics[32], a finite-element solver that can solve arbitrary sets of equations and boundary conditions. The calculation of carrier concentration distributions and electromagnetic wave scattering can then be carried out using the same mesh within a single software package.

The simulation of the electronic properties of the semiconductor layer can be accomplished with some generality. The electron and hole concentrations in a semiconductor subject to an external quasistatic electric field (or voltage) obey the well-known[31] drift-diffusion equations,

$$\frac{\partial n_e}{\partial t} + \frac{1}{-e}\nabla \cdot \mathbf{j}_e = -R_{SRH}$$
$$\frac{\partial n_h}{\partial t} + \frac{1}{e}\nabla \cdot \mathbf{j}_h = -R_{SRH} \qquad (4)$$

coupled with the electrostatic Poisson equation,

$$\nabla \cdot \mathbf{D} = -\nabla(\varepsilon \nabla \phi) = e(n_h - n_e + N_d), \tag{5}$$

where $\phi$ is the electrostatic potential. The current densities due to electrons and holes in Eqs. (4) include the drift and diffusion terms:

$$\begin{aligned} \mathbf{j}_e &= -e n_e \mu_e \nabla \phi + e D_e \nabla n_e, \\ \mathbf{j}_h &= -e n_h \mu_h \nabla \phi - e D_h \nabla n_h, \end{aligned} \tag{6}$$

where $\mu_{e,h}$ are the carrier mobilities and $D_{e,h}$ their diffusivities. In the limit where Fermi-Dirac distributions can be approximated by Boltzmann distributions, i.e., $n_e \propto \exp[-(-e)\phi/(k_B T)]$ and $n_h \propto \exp[-e\phi/(k_B T)]$, the mobilities and diffusivities are related by the Einstein equations $D_{e,h} = \mu_{e,h} k_B T / e$. For the simulations presented here, we use[35] $\mu_e = 0.1 \, m^2/(V \cdot s)$ and $\mu_h = 0.05 \, m^2/(V \cdot s)$.

The electron-hole recombination rate on the r.h.s. of Eqs. (4) is approximated using the Shockley-Read-Hall formula[31],

$$R_{SRH} = \frac{n_e n_h - n_i^2}{\tau_h (n_e + n_{e1}) + \tau_e (n_h + n_{h1})}, \tag{7}$$

where $n_i$ is the temperature-dependent intrinsic carrier concentration, $\tau_{e,h}$ are the carrier lifetimes, and $n_{e1}$ and $n_{h1}$ are parameters related to the trap energy level. If the trap energy level is located at the middle of the band gap, $n_{e1} \approx n_{h1} \approx n_i$, which is the approximation assumed in the calculations presented here. For our calculations, we assume[31] $\tau_e = \tau_h = 10^{-7} \, s$ and $n_i = 1.45 \times 10^{10} \, cm^{-3}$. The combined doping concentration in the right hand side of Eq. (5) includes both the *n*- (donor) and *p*-type (acceptor) dopant concentrations: $N_d = N_D - N_A$. In our specific example, we consider *n*-type doped Si with doping concentration $N_d = N_D = 2.3 \times 10^{15} \, cm^{-3}$, the value estimated from the DC conductivity of the Si samples.

The set of three partial differential equations provided in Eqs. (4-5) can be solved numerically by imposing a set of boundary conditions over the computational domain. The *insulation boundary* has no surface charge and no current passing through it. We thus apply the boundary conditions

$$\hat{n} \cdot \mathbf{D} = 0$$
$$\hat{n} \cdot \mathbf{j}_e = \hat{n} \cdot \mathbf{j}_h = 0 \tag{8}$$

at the insulation boundary.

As described above, Schottky contacts are characterized by a barrier potential $\phi_B$ (Fig. 4a), as well as by electron and hole currents that flow whenever the charge carriers are not in thermal equilibrium. The set of boundary conditions at a Schottky contact, then, consists of

$$\phi = V_a - \phi_B$$
$$\hat{n} \cdot \mathbf{j}_e = ev_e \left[ n_e - n_e^{eq}(\phi) \right]$$
$$\hat{n} \cdot \mathbf{j}_h = ev_h \left[ n_h - n_h^{eq}(\phi) \right], \tag{9}$$

where $\hat{n}$ is the unit normal to the boundary,

$$n_e^{eq}(\phi) = n_i \exp\left\{ \frac{\phi + \chi_{semi} + E_g/2}{k_B T / e} \right\} \tag{10}$$

and

$$n_h^{eq}(\phi) = n_i^2 / n_e^{eq}(\phi) = n_i \exp\left\{ -\frac{\phi + \chi_{semi} + E_g/2}{k_B T / e} \right\} \tag{11}$$

are the equilibrium electron and hole concentrations, and $v_e$ and $v_h$ are phenomenological parameters; we use[37-38] $v_e = 2.207 \times 10^4 \, m/s$ and $v_h = 1.62 \times 10^4 \, m/s$.

Assuming that the voltage bias is stationary and the electron and hole currents have reached their steady state, spatial distributions of $n_e$ and $n_h$ can be found by neglecting the time derivative and solving Eqs. (4-5). These equations form a strongly coupled, strongly nonlinear set, which we solve using COMSOL Multiphysics[32]. Because the problem is strongly nonlinear, both due to the partial differential equations themselves and the boundary conditions, an iterative nonlinear solver is used, which requires a good initial guess for the potential and concentration profiles. The initial guess can be obtained at zero voltage bias by making an assumption of thermal equilibrium and using Eqs. (10-11) in combination with the electrostatic Poisson equation (5).

Negative voltage bias at the Schottky contacts on *n*-type doped semiconductors results in an increase of the depletion region(s) width. Outside of the depletion region, the electron concentration is virtually unchanged, as in the absence of a bias. We assume here that the *n*-type dopant concentration is uniform in the entire semiconductor layer. The transition region between the depleted and undepleted regions has a thickness of roughly 200 nm, as can be seen in the simulated carrier density profiles in Fig. 5. Within this extremely thin layer, electron concentrations change by 4-9 orders of magnitude; the finite element mesh needed to accurately resolve such a huge gradient must be extremely fine. From numerical experiments, we determine that fast convergence towards the correct solution requires a mesh size in the transition region no larger than 5 nm.

As an initial test of the approach, we first model a 1D contact and compare with the analytical result of Eq. 2. The comparison, shown in Fig. 4b, reveals virtually exact agreement of the depletion depth as a function of bias voltage for the 1D contact. For the simulated carrier profile, the depletion depth is defined at the point where the carrier density is half way between fully depleted and non-depleted densities.

Once the electron and hole distributions in the semiconductor are calculated, the spatially inhomogeneous and dispersive dielectric function of silicon is calculated using Eq. (1), and the full-wave, frequency domain Maxwell's equations are solved for a single-layer, two-dimensionally periodic metamaterial slab. Both the semiconductor physics and the electromagnetic simulations are performed on the same geometry using the same finite element mesh; however, our algorithm allows using different meshes for the discretization of these different equations, which is beneficial since the mesh used to solve the semiconductor problem is much finer than required for solving the electromagnetic wave problem. The electromagnetic excitation in our simulations is a monochromatic, linearly polarized plane wave. The effect of the high-frequency field on the conductivity of the semiconductor layer is neglected, which is a valid approximation for low enough incident intensities at which the metamaterial does not have a measurable nonlinear response. This assumption is confirmed at the time of measurement by observing that the measured reflection and transmission spectra are not dependent upon the power level of the incident field.

## 5. Metamaterial characterization in the W-band

To characterize the ELC sample, a vector network analyzer (VNA, Agilent Technologies,

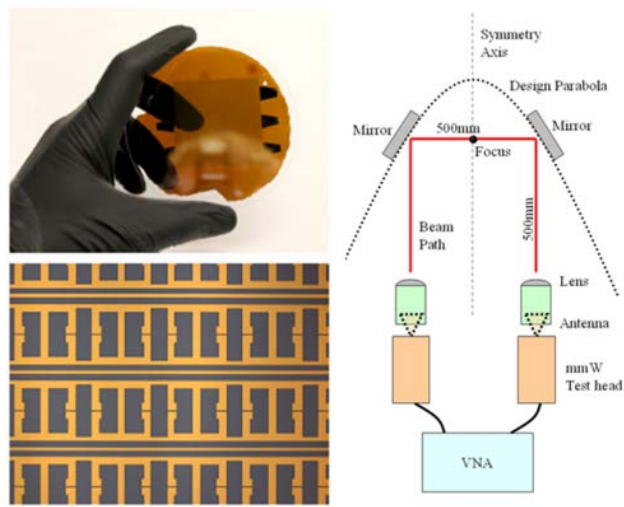

**Figure 6 | Sample and experimental configuration.** Left, top: Large area metamaterial sample on a Pyrex substrate, as fabricated. Left, bottom: Optical microphotograph showing the ELCs and bias lines. Right: The experimental configuration used to measure the sample transmission and reflection coefficients (S-parameters).

PNA-E8361A) was used to generate radiation over the 60-90 GHz frequency range (W-band). The measurement configuration, shown in Fig. 6, consists of two standard gain rectangular horn antennas, each with a dielectric lens used to produce a collimated beam, and two focusing mirrors placed in each beam path that direct energy to the metamaterial sample. The lens and the antenna are placed to give an f-stop of f/0.7. The mirrors are cut from a design parabola 250 mm off-center from the axis of symmetry. The parabola has a focus of 250 mm. The mirrors are made from aluminum blocks, ground and polished. The metamaterial sample to be tested is placed in a sample holder at the focus of the design parabola where it is illuminated by the W-band radiation. The beam forms an oval cross section at the focal point, with a measured major axis of 3.8 cm and minor axis of 3.1 cm.

A standard power supply is used to apply a variable bias voltage to the sample, between 0-25 V. At a number of applied voltages, the reflectance and transmittance of the sample are measured, and serve as a means of confirming the predicted behavior of the Schottky contacts. To reduce the impact of voltage standing wave ratio (VSWR) in the power reflected from the sample ($S_{11}$) caused by multiple reflections between the lenses, mirrors and other elements, time gating was employed using the time domain option on the VNA, as described in Agilent Technologies documentation[39-41]. Before measuring the sample, a GRL (gated-reflection-line) calibration is performed on the system in conjunction with the time domain option. The wave propagation path length is chosen such that the quasi-optical elements are distant from one another to ensure a quality GRL calibration.

Time domain analysis and time gating are useful techniques at gigahertz frequencies to improve measurement quality[39]. Network analyzers have the ability to transform the frequency domain measurement into the time domain via Fourier transform. Another useful property of time domain analysis is to design a frequency domain filter with information from the time domain so that unwanted discontinuities are removed from the measurement. If, for example, there is a chain of components in the device under test (DUT), and a measurement should be performed on only one of the components, then the rest can be time gated out. While this method is named 'time gating', the time domain is used to only locate the correct DUT and design the frequency domain filter with this information. The filter is applied directly to the frequency domain measurement, to avoid the issues that would arise from applying an inverse Fourier transform, gating, and then Fourier transforming back into the frequency domain.

The biggest issue with time gating is masking. Masking occurs when a large reflection before the discontinuity of interest is time gated out of the measurement. A large portion of the signal power generated by the network analyzer never makes it to the discontinuity of interest because of the masked large reflection. The network analyzer typically does not account for this and, as a result, the discontinuity of interest looks as though it is lossier than it really is. In our setup, however, all of the reflection points before the metamaterial wafer are unchanged between measurements. This repeatable error can be calibrated out using two measurements, one using a perfect transmission device (air), and one using a perfect reflector device (a metal plate). The rest of unwanted reflections in the free-space measurement system can be absorbed into other calibration terms in the network analyzer. This calibration technique is known as gated-reflection-line (GRL)[42]. With this calibration technique, the only worry with masking is a decrease of dynamic range in the measurement system due to unwanted reflections removing signal power from the system. A well designed system can minimize the reduction of the dynamic range.

A secondary issue arising from time gating is the smoothing of responses over frequency due to the roll-off of the frequency domain filter. The smoothing can be reduced by making the filter roll-off more aggressive, but this can reintroduce unwanted reflections that are nearby the DUT if they reside in the sidelobes. Most of the unwanted reflections in our system come from the network analyzer-antenna interface and the antenna-air interface. To allow for aggressive filter design, the path to the DUT from the antennas was made long to space out the responses of the antenna and DUT in time.

The described apparatus and technique was also used to characterize the Pyrex substrates, whose dielectric constant is later used in numerical models. To retrieve the complex-valued refractive index, we have measured the magnitude and phase of the $S_{11}$ and $S_{21}$ parameters (reflection and transmission coefficients) at normal incidence, and applied the index retrieval technique based on the inversion of Fresnel-Airy formulas, sometimes known as the Nicolson-Ross-Weir method[41,43-45]. This yielded real part of the dielectric constant ($\varepsilon'$) almost flat in the entire W-band, with values in the 4.2-4.3 range, consistent with the prior experimental studies of

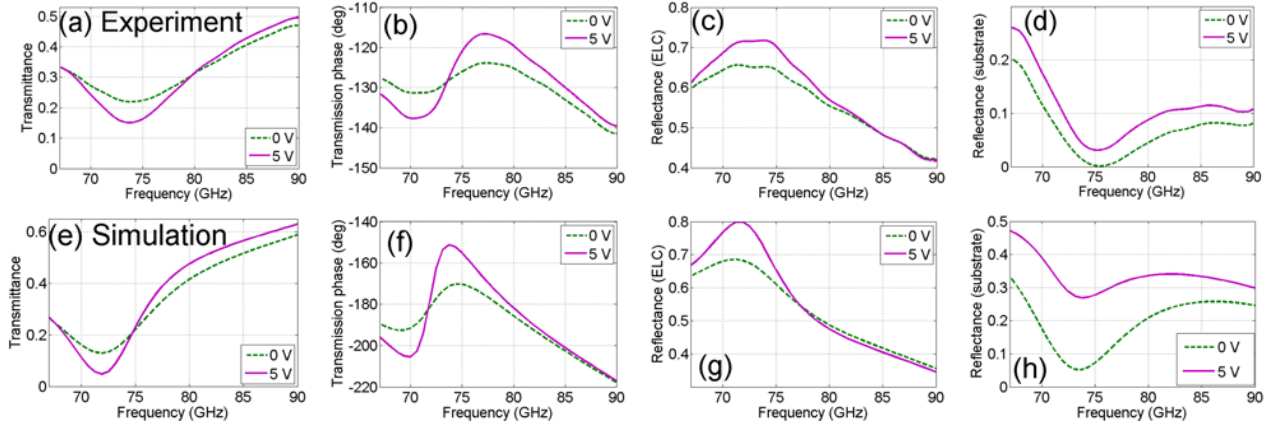

**Figure 7 | Simulated and measured transmittance and reflectance data for the ELC metamaterial.** (a) Transmittance ($T = |S_{21}|^2$), (b) phase of the transmission coefficient, (c) reflectance from the ELC side ($R_1 = |S_{11}|^2$) and (d) reflectance from the Pyrex substrate side ($R_2 = |S_{22}|^2$) of the sample. (e), (f), (g) and (h) are the same quantities, as simulated.

Pyrex glass at millimeter wavelengths[46-47]. The dielectric loss tangent was fluctuating in the range 0-0.04, with an average value of about 0.015. Notably, another retrieval method provided by Agilent software and referred to as "Transmission Epsilon Fast"[41], based solely on transmittance fitting and the assumption of unit permeability ($\mu=1$), gave $\varepsilon'$ in the range 4.35-4.7, with an average value about 4.45. Although we use $\varepsilon_{Py} \approx 4.2$, the spread in experimental measurements creates an uncertainty in the $\varepsilon$ of the substrate, which affects our predictions of reflectance coefficients and the metamaterial resonance frequency; the latter is particularly sensitive to the permittivity of the substrate.

Because the metamaterial-on-Pyrex sample does not have reflection symmetry about the axis of propagation, the reflection coefficient is not the same for waves incident from the two network analyzer ports. The reflectance was therefore measured in both directions ($S_{11}$ and $S_{22}$) for the full characterization. The measured data are shown in Figs. 7a, b, c and d. While Fig. 7 shows only the data and simulations for 0-5 V bias, the agreement is very similar for all bias values in the 0-6.5 V. Our modeling approach cannot be used to accurately predict concentration profiles beyond 6.5 V, where depletion depth becomes greater than the Si device layer (2 μm). Good agreement between simulation and measurement can be seen from the comparison of the spectral features of the scattering (S-) parameters near the metamaterial resonance as a function of applied bias. As can be seen in Fig. 7, the resonance increases in strength as the reverse bias voltage is increased, causing transmission reduction near the resonance frequency. Qualitatively, the S-parameters are in excellent agreement with the simulations, and are also quite close in terms of quantitative agreement. The most important figure of merit of this tunable metamaterial – maximum variation of the transmission coefficient near the resonance (about 8% per 5 V bias) – is in excellent agreement between the simulation and experiment.

While the bias differentials of S-parameters as obtained from the simulations are in fair agreement with experimental measurements, the absolute values of S-parameters are somewhat different. This is to be expected due to the variability and uncertainty of the Si device layer thickness, which is expected to vary by up to ±25% across the device area. Silicon thickness variation causes a proportional fluctuation of the absorption coefficient per unit cell. Additionally, since the damping rate of the metamaterial resonators depends on (a) Si layer thickness and (b) the doping concentration (which is uncertain by at least 10-15%), fluctuations of (a) and (b) can cause noticeable variations in the real part of effective refractive index of this metamaterial. It is therefore possible that a fraction of the beam used in S-parameter measurements was deflected by diffraction through a variable-index metamaterial layer and not collected at the receiver port. These variations and uncertainties explain the discrepancy between simulated and measured S-parameters, which reaches 20% in reflectance at the high-frequency end of the measurement band.

As described above, the simulation results were obtained by first performing the electrostatic analysis on the Si-ELC geometry, and subsequently using the computed carrier concentrations to determine the electromagnetic S-parameters. The semiconductor parameters for Si were used as input, with no adjustable parameters. The agreement obtained in Fig. 7 indicates that the operation of semiconductor metamaterials integrated with *mm*-wave devices, can be predicted using known analysis techniques with the accuracy that is only limited by the accuracy of determination of a few material constants, such as the semiconductor doping levels and the dielectric constant of the substrate.

### 6. Discussion and conclusions

The tuning of a single metamaterial layer can bring considerable functionality and reconfigurability to the aperture of an antenna or imaging system, especially if each element becomes individually addressable. Dynamically reconfigurable diffractive or holographic optics—in which far field patterns are formed by controlling the amplitude and phase advance of a field across an aperture—could be implemented using the approaches illustrated here.

In addition to the planar aperture applications, semiconductor tuning could also be used to form reconfigurable bulk metamaterials, which might be useful for gradient-index or transformation optical devices. It is of interest to examine the range of equivalent, bulk constitutive parameters achieved by the tuning demonstrated in the present device. While actual gradient-index or transformation optical devices based on such a reconfigurable metamaterial are more likely than not to consist of periodically repeated layers, we choose to characterize only a single layer metamaterial, with the goal of evaluating the degree to which its effective medium parameters can be tuned as a function of applied voltage. Modeling a single layer of a metamaterial, which induces a phase delay not in excess of 180°, makes it easy to apply the standard S-parameter retrieval technique, which would otherwise be difficult to apply. Multilayer metamaterials and their effective medium parameters should be characterized with different techniques[48-49], which are beyond the scope of this paper.

Using either the measurement or the simulation data, in principle, one can apply standard retrieval methods that use the S-parameter data to obtain the effective constitutive parameters for the ELC metamaterial sample[50-51]. In our case, however, the Si-ELC metamaterial was fabricated on an optically thick (non-subwavelength) Pyrex substrate, which rendered homogenization of the metamaterial-substrate sandwich impossible. We thus perform homogenization of a simulated structure, where one can assume the Pyrex substrate to be removed (Figure 8). The overall effect of Pyrex on the metamaterial appears to be mostly a resonance redshift: the resonance occurs at 72 GHz with Pyrex and at 94 GHz without it. Since the S-parameters generally show good agreement between our simulations and measurements in the case of Pyrex-mounted metamaterial (Fig. 7), we expect that a simulation of a free-standing Si-ELC layer would also be in reasonable agreement with a measurement, if it could be conducted.

The retrieved effective permittivity is shown in Fig. 8, as a function of the applied voltage. As is typical for ELC structures, the permittivity reaches extremely large values near the resonance. The magnetic permeability (not shown) is close to unity in the entire frequency range, showing only a minor feature near the electric resonance frequency, a phenomenon well known as anti-resonance.[52] Increasing the bias voltage causes the local concentration of carriers to decrease, due to the increase in width of the depletion region. The effect is that the resonance is less damped, and the range of permittivity values increases substantially. Again, for this simulation, the unit cell in the propagation direction comprises the 2 μm thick silicon layer and the 230 nm metallic ELC layer, making the assumed metamaterial density extremely large. For this reason, effective permittivity reaches extreme values that may be of use in certain scenarios; the large response, however, is accompanied by large damping and thus large effective loss tangents. In addition, the permittivity distribution is highly anisotropic: the permittivity in directions perpendicular to the stacking would be significantly lower.

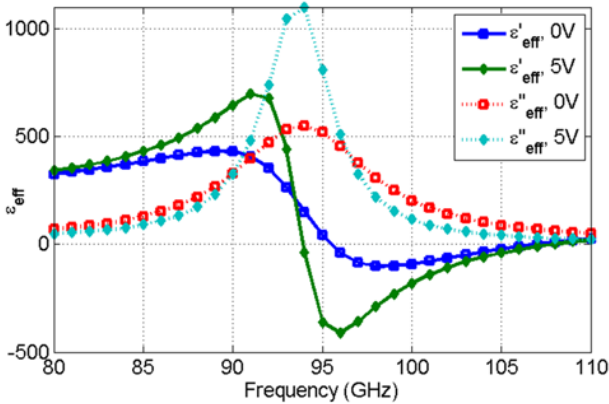

**Figure 8 | Effective permittivity for a free-standing single layer of ELC metamaterial.** Simulated effective relative permittivity corresponding to the orientation of the electric field in the ELC plane and across the ELC gap. Pyrex substrate is removed.

We have presented a tunable metamaterial based on interfacing metamaterial elements with *n*-doped thin film Si bonded to a Pyrex wafer. The use of thin film Si is of particular interest, since so many well-established fabrication processes have been developed for Si and the substrate losses due to thick semiconductors can be minimized by bonding to a low loss substrate. In any production quantity, Si metamaterial devices would be inexpensive and robust, as they leverage a mature industry. Though the carrier mobility in silicon is not as large as in other semiconductors, such as GaAs, the expected switching speeds for Si are appropriate for a variety of beam-forming and imaging applications. In particular, single pixel imaging devices that rely on the modulation of the transmission or reflection of a collection of elements could be readily formed using a voltage controlled metamaterial such as that shown here[53].

Given the potential for active metamaterial-semiconductor devices,[54-56] the combined modeling of semiconductor physics and electromagnetic scattering becomes attractive. The results found here demonstrate that quantitative predictions of a complete, dynamically controlled metamaterial device can be achieved. The length scales for the two different simulations proved to be challenging for the work presented here, but we anticipate that continued development of the modeling tools will result in much greater efficiency and additional methods to handle the multiscale nature of the problem. From the promising results in this work, it is likely that far more complicated integrated semiconductor devices can be modeled and realized.

**Acknowledgements**

This work was supported by Toyota Motor Engineering and Manufacturing North America, and partially supported by the Air Force Office of Scientific Research (Grant No. FA9550-09-1-0562). The authors are thankful to Erik Danielsson (COMSOL, Sweden) for a useful discussion on semiconductor physics modeling.

*To whom the correspondence should be addressed:  yaroslav.urzhumov@duke.edu

**References**

[1] N. Engheta and R. W. Ziolkowski, *Electromagnetic Metamaterials: Physics and Engineering Explorations* (Wiley-IEEE Press, 2006).

[2] L. Solymar and E. Shamonina, *Waves in Metamaterials* (Oxford Univ. Press, 2009).

[3] R. Marques, F. Martin, and M. Sorolla, *Metamaterials with Negative Parameters: Theory, Design and Microwave Applications* (Wiley, 2008).

[4] R. Walser, in *Proceedings of SPIE*, Vol. 4467 (2001) p. 1.

[5] D. Smith, W. Padilla, D. Vier, S. Nemat-Nasser, and S. Schultz, Phys. Rev. Lett. **84**, 4184 (2000).

[6] S. Zhang, W. Fan, N. Panoiu, K. Malloy, R. Osgood, and S. Brueck, Phys. Rev. Lett. **95**, 137404 (2005).

[7] V. Shalaev, W. Cai, U. Chettiar, H. Yuan, A. Sarychev, V. Drachev, and A. Kildishev, Opt. Lett. **30**, 3356 (2005).

[8] D. Schurig, J. J. Mock, B. J. Justice, S. A. Cummer, J. B. Pendry, A. F. Starr, and D. R. Smith, Science **314**, 977 (2006).

[9] J. Valentine, J. Li, T. Zentgraf, G. Bartal, and X. Zhang, Nature Mater. **8**, 568 (2009).

[10] T. Ergin, N. Stenger, P. Brenner, J. Pendry, and M. Wegener, Science **328**, 337 (2010).

[11] L. Kang, Q. Zhao, H. Zhao, and J. Zhou, Optics Express **16**, 8825 (2008).

[12] H.-T. Chen, H. Yang, R. Singh, J. F. O'Hara, A. K. Azad, S. A. Trugman, Q. X. Jia, and A. J. Taylor, Phys. Rev. Lett. **105**, 247402 (2010).

[13] T. Driscoll, S. Palit, M. Qazilbash, M. Brehm, F. Keilmann, B. Chae, S. Yun, H. Kim, S. Cho, and N. Jokerst, Appl. Phys. Lett. **93**, 024101 (2008).

[14] M. Dicken, K. Aydin, I. Pryce, L. Sweatlock, E. Boyd, S. Walavalkar, J. Ma, and H. Atwater, Opt. Express **17**, 18330 (2009).

[15] R. Singh, A. Azad, Q. Jia, A. Taylor, and H. Chen, Optics Lett. **36**, 1230 (2011).

[16] L. Ju, B. Geng, J. Horng, C. Girit, M. Martin, Z. Hao, H. Bechtel, X. Liang, A. Zettl, Y. Shen and Feng Wang, Nature Nanotechnol. **6**, 630 (2011).

[17] T. Kasirga, Y. Ertas, and M. Bayindir, Applied Physics Letters **95**, 214102 (2009).

[18] I. Pryce, K. Aydin, Y. Kelaita, R. Briggs, and H. Atwater, Nano Lett. **10**(10), 4222 (2010).

[19] J. Ou, E. Plum, L. Jiang, and N. Zheludev, Nano Lett. **11**(5), 2142 (2011).

[20] H. Tao, A. Strikwerda, K. Fan, W. Padilla, X. Zhang, and R. Averitt, Phys. Rev. Lett **103**, 147401 (2009).

[21] W. Padilla, A. Taylor, C. Highstrete, M. Lee, and R. Averitt, Phys. Rev. Lett. **96**, 107401 (2006).

[22] C. Lee, P. Mak, and A. DeFonzo, Quantum Electronics, IEEE Journal of **16**, 277 (1980).

[23] J. Li, Optics & Laser Technology **43**, 102 (2011).

[24] A. Degiron, J. Mock, and D. Smith, Optics Express **15**, 1115 (2007).

[25] H. Chen, J. O'Hara, A. Azad, A. Taylor, R. Averitt, D. Shrekenhamer, and W. Padilla, Nature Photon. **2**, 295 (2008).

[26] H. Chen, W. Padilla, J. Zide, A. Gossard, A. Taylor, and R. Averitt, Nature **444**, 597 (2006).

[27] H. Chen, W. Padilla, M. Cich, A. Azad, R. Averitt, and A. Taylor, Nature Photon. **3**, 148 (2009).

[28] H. Chen, S. Palit, T. Tyler, C. Bingham, J. Zide, J. OHara, D. Smith, A. Gossard, R. Averitt, W. Padilla, *et al.*, Appl. Phys. Lett. **93**, 091117 (2008).

[29] D. Shrekenhamer, S. Rout, A. Strikwerda, C. Bingham, R. Averitt, S. Sonkusale, and W. Padilla, Opt. Express **19**, 9968 (2011).

[30] D. Schurig, J. Mock, and D. Smith, Appl. Phys. Lett. **88**, 041109 (2006).

[31] S. M. Sze and K. K. Ng, *Physics of Semiconductor Devices*, 3rd ed. (John Wiley & Sons, New York, 2007).

[32] *COMSOL Multiphysics User's Guide, Version 4.1*, COMSOL AB, Burlington, Mass., USA (2010).

[33] H. Tao, A. C. Strikwerda, K. Fan, W. J. Padilla, X. Zhang, and R. D. Averitt, Phys. Rev. Lett. **103**, 147401 (2009).

[34] C. H. Lee, P. S. Mak, and A. P. DeFonzo, IEEE Journal of Quantum Electronics **16**, 277 (1980).

[35] M. A. Green, J. Appl. Phys. **67**, 2944 (1990).

[36] H. Muta, Japanese J. Appl. Phys. **17**, 1089 (1978).

[37] N. Toyama, T. Takahashi, H. Murakami, and H. Koriyama, Appl. Phys. Lett. **46**, 557 (1985).

[38] N. Toyama, T. Takahashi, H. Murakami, and H. Koriyama, J. Appl. Phys. **63**, 2720 (1988).

[39] *Specifying Calibration Standards and Kits for Agilent Vector Network Analyzers, Application Note 1287-11*, Agilent Technologies.

[40] *PNA Microwave Network Analyzers - Banded Millimeter-Wave Measurements with the PNA, Application Note 1408-15*, Agilent Technologies.

[41] *Materials Measurement Software - Technical Overview, Technical Note 85071E*, Agilent Technologies.

[42] P. Bartley and S. Begley, in *Instrumentation and Measurement Technology Conference, 2005. IMTC 2005 - Proceedings of the IEEE*, Vol. 1, 372–375 (2005).

[43] A. M. Nicolson and G. F. Ross, Instrumentation and Measurement, IEEE Transactions on **19**, 377 (1970).

[44] W. Weir, Proceedings of the IEEE **62**, 33 (1974).

[45] D. K. Ghodgaonkar, V. V. Varadan, and V. K. Varadan, IEEE Transactions on Instrumentation and Measurement **37**, 789 (1989).

[46] J. W. Lamb, International Journal of Infrared and Millimeter Waves **17**, 1997 (1996).

[47] M. Halpern, H. P. Gush, E. Wishnow, and V. D. Cosmo, Appl. Opt. **25**, 565 (1986).

[48] M. Davanco, Y. Urzhumov, and G. Shvets, Opt. Express **15**, 9681 (2007).

[49] C. Fietz and G. Shvets, Physica B: Condensed Matter **405**, 2930 (2010).

[50] D. R. Smith, S. Schultz, P. Markoŝ, and C. M. Soukoulis, Phys. Rev. B **65**, 195104 (2002).

[51] D. Smith, D. Vier, T. Koschny, and C. Soukoulis, Phys. Rev. E **71**, 036617 (2005).

[52] T. Koschny, P. Markoš, D. R. Smith, and C. M. Soukoulis, Phys. Rev. E **68**, 065602 (2003).

[53] D. Mittleman, M. Gupta, R. Neelamani, R. Baraniuk, J. Rudd, and M. Koch, Appl. Phys. B: Lasers and Optics **68**, 1085 (1999).

[54] M. Knight, H. Sobhani, P. Nordlander, and N. Halas, Science **332**, 702 (2011).

[55] N. Meinzer, M. Ruther, S. Linden, C. M. Soukoulis, G. Khitrova, J. Hendrickson, J. D. Olitzky, H. M. Gibbs, and M. Wegener, Opt. Express **18**, 24140 (2010).

[56] S. Schwaiger, M. Klingbeil, J. Kerbst, A. Rottler, R. Costa, A. Koitmäe, M. Bröll, C. Heyn, Y. Stark, D. Heitmann, *et al.*, Phys. Rev. B **84**, 155325 (2011).